\pdfoutput=1
\documentclass[10pt]{article}

\usepackage{booktabs}
\usepackage{threeparttable}
\usepackage{color}
\usepackage{bbm}
\usepackage{epsfig}
\usepackage{amssymb,amsmath,amscd, mdwmath}
\usepackage{graphicx,cite,url}
\usepackage{algorithmic}
\usepackage{float}
\usepackage{amsfonts}
\usepackage{syntonly}
\usepackage{multirow}
\usepackage{graphicx}
\usepackage{subfig}
\usepackage{tabularx}
\usepackage{setspace}
\usepackage{stfloats}
\usepackage{amsthm}
\usepackage[labelfont=bf,labelsep=period,justification=raggedright]{caption}

\makeatletter
\renewcommand{\@biblabel}[1]{\quad#1.}
\makeatother

\date{}

\begin{document}

\begin{flushleft}
{\Large
\textbf{Assessing   the level of merging  errors for coauthorship data: a Bayesian model   }
}
\\
Zheng Xie$^{1, * }$
\\
\bf{1}   College of Liberal Arts and Sciences,   National University of Defense Technology, Changsha,   China  
\\ $^*$ xiezheng81@nudt.edu.cn.
 \end{flushleft}

\section*{Abstract}

Robust analysis of coauthorship networks is based on high quality data. However, ground-truth data are usually unavailable. Empirical data suffer several types of errors,
  a typical one of which is called merging error,
identifying different persons as one entity.
   Specific features of authors have been used to reduce  these errors. We proposed a Bayesian model to calculate the information of any given features  of authors. Based on the   features, the model can be utilized to calculate the rate of merging errors    for entities. Therefore, the model helps to find  informative features for detecting heavily compromised  entities . It has potential contributions to improving the quality of empirical data.




\section*{Introduction}

 Studies of coauthorship networks with large scale
  provide a bird-eye view of collaboration patterns in
 scientific society, and have become an important topic of social sciences\cite{Adams,Glanzel1,Uzzi}.
Ambiguities  exist  in the collection  of coauthorship data,  which   manifest  themselves in two ways: one
 person   is identified as two or more  entities  (splitting error);  two or more persons   are
identified as one entity (merging error). Here     persons      refer to  authors, and
   entities   refer  to   nodes of coauthorship networks\cite{Smalheiser}.
 These errors  make the collected  data deviate from ground truth.
Merging
errors   deflate  the number of persons, the average shortest path,  and the global clustering
coefficient\cite{Kim1,Milojevic2,WangDJ}. These errors also   inflate  the average number of the papers  per author, the average number of the collaborators   per
author, and the size of giant component.

 Analysis drawn on imperfect data is at risk.  However,  in data-intensive   research, ground-truth  data are often unavailable.
In social network analysis, there exist extensive research on  detecting and reducing   data errors\cite{Oliveira}, especially   on missing  and spurious connections\cite{Guimera,Han,Xie0}.
 In the case of coauthorship networks, such research   is called name disambiguation.
For the authors of different papers with the same name, it is reasonable to assume that the more similar (such coauthors, email, etc.) these authors' features are, the more likely they are of the same person. Therefore, a range of name disambiguation methods have been provided to reduce  the merging and splitting errors, based on the features extracted from the contents of publications (such as email address, title, abstract, and references), from the web information of authors, and so on\cite{ Angelo,Ferreira, Tang}. Those methods have been proved to be effective in specific cases\cite{Franceschet,Ley,Onodera,Torvik,Zhao}. Utilizing the features of authors efficiently is still a challenge and needs further research\cite{Hussain}.
Here we do not consider the specific techniques of reducing  errors, but want to know the merging error level of empirical data.
Whether the level  is severe enough to warrant data reliability? Which  data are most heavily compromised?


In the view of Bayesians, inference from data is the revision of a given opinion in the light of relevant new information\cite{Edwards}. Therefore, we proposed a Bayesian model for utilizing observed author features to revise the opinion reflected by empirical data. Our model is based on the Bayesian model provided by Newman, which is designed to infer network structures\cite{Newman9}. The novel aspect here is that we utilized the prior knowledge of error rates. This  is inspired by the Bayesian model provided by Butts, which is designed to infer the errors of networks\cite{Butts}. The different aspect from Butts' model is that we removed the hyperparameters of the distribution of ground-truth data.
 Our model can be utilized to calculate the information of authors' features. When knowing which features are informative, we can use them to calculate the rate of merging errors over entities effectively, and then find those entities heavily compromised. Therefore, the model has the ability to assess the error level of coauthorship networks,
and to
estimate   the number of nodes  for those networks.
  We demonstrated model  functions by using the names of coauthors, a feature used in many name disambiguation methods.

This report is organized as follows:  the model is described in
Section  2;   its demonstration is shown
   Section  3;  the discussion and conclusions are drawn in Section 4.

\section*{The  model}

 Consider $N_l$ authors to be of  an  entity $l$.
  Denote
the ground-truth relationship  by  $A_{ij}$ for all author pair $(i,j)$,
where $A_{ij}=1$ means the  pair are actually one person, and $A_{ij}=0$ means not.
 That is, the probability of  authors $i$ and $j$ being the same  person  depends only on $A_{ij}$.  This dependence can be
parameterized by two quantities: the false negative rate $e^-$, and  the false positive rate $e^+$ that is merging error rate.
 Note that $e^-$  is not splitting error rate, because it does not address the case that a person is regarded  as several entities.

Assume  that
the prior distribution of $A_{ij}$ is determined via a  parameter
  $\rho\in[0,1]$, the specific form of which adopted  here  is   $A_{ij}\sim \rho^{A_{ij}}(1-\rho)^{1-A_{ij}}  $.
   Then the likelihood function of $\rho$ is
    \begin{align}\label{eq1}L(\rho)&=\prod_{i<j}\rho^{A_{ij}}(1-\rho)^{1-A_{ij}}  . \end{align}
   Maximizing it respect to $\rho$ gives rise to
     \begin{align}\label{eq2}\rho&=\frac{2\sum_{i<j}A_{ij}}{N_l(N_l-1)}. \end{align}

     Assume  that  the error rates  $e^+$ and $e^-$ are drawn
independently from two Beta distributions, namely $e^+\sim  \mathrm{Beta}(\alpha^+, \beta^+)$ and $e^- \sim \mathrm{Beta}(\alpha^-, \beta^-)$.
The choice of the parameters of these distributions  reflects the
  prior knowledge regarding the errors of data.
 The  density function of  a Beta distribution $\mathrm{Beta}(\alpha, \beta)$ is
  $\mathrm{Beta}(x|\alpha, \beta)=\Gamma(\alpha+ \beta)x^{\alpha-1}(1-x)^{\beta-1}/(\Gamma(\alpha)\Gamma(\beta))$, where $x\in [0,1]$.

 Specify a real value $N_{ij}$ to  quantify  any observed   features (such as coauthor names, affiliation,
      email, and so on) of author pair  $(i,j)$,
 and a real value $E_{ij}$ to  quantify the    similarity  between their features.
  So
$E_{ij}/ N_{ij}\in[0,1]$ expresses the similar level between $i$ and $j$.
Assume  that the probability of observing
the   features and similarity
(quantified  as $E_{ij}$ and $N_{ij}$) conditional on    $A_{ij}$, $e^+$ and   $e^-$ is the following Bernoulli
mixture
\begin{align}\label{eq3}p(E_{ij},N_{ij}|A_{ij},  e^+, e^- )=  B(E_{ij},N_{ij}|1-e^-)^{A_{ij}} \times B(E_{ij},N_{ij}|e^+) ^{(1-A_{ij}) },
 \end{align}
where $B(E_{ij},N_{ij}|x)$ is the   density $\mathrm{Beta}(x|E_{ij}+1, N_{ij}- E_{ij}+1)$ for $x\in\{e^+,  1-e^-  \}$.

\begin{figure*}[h]
\centering
 \includegraphics[height=3.2  in,width=4.5  in,angle=0]{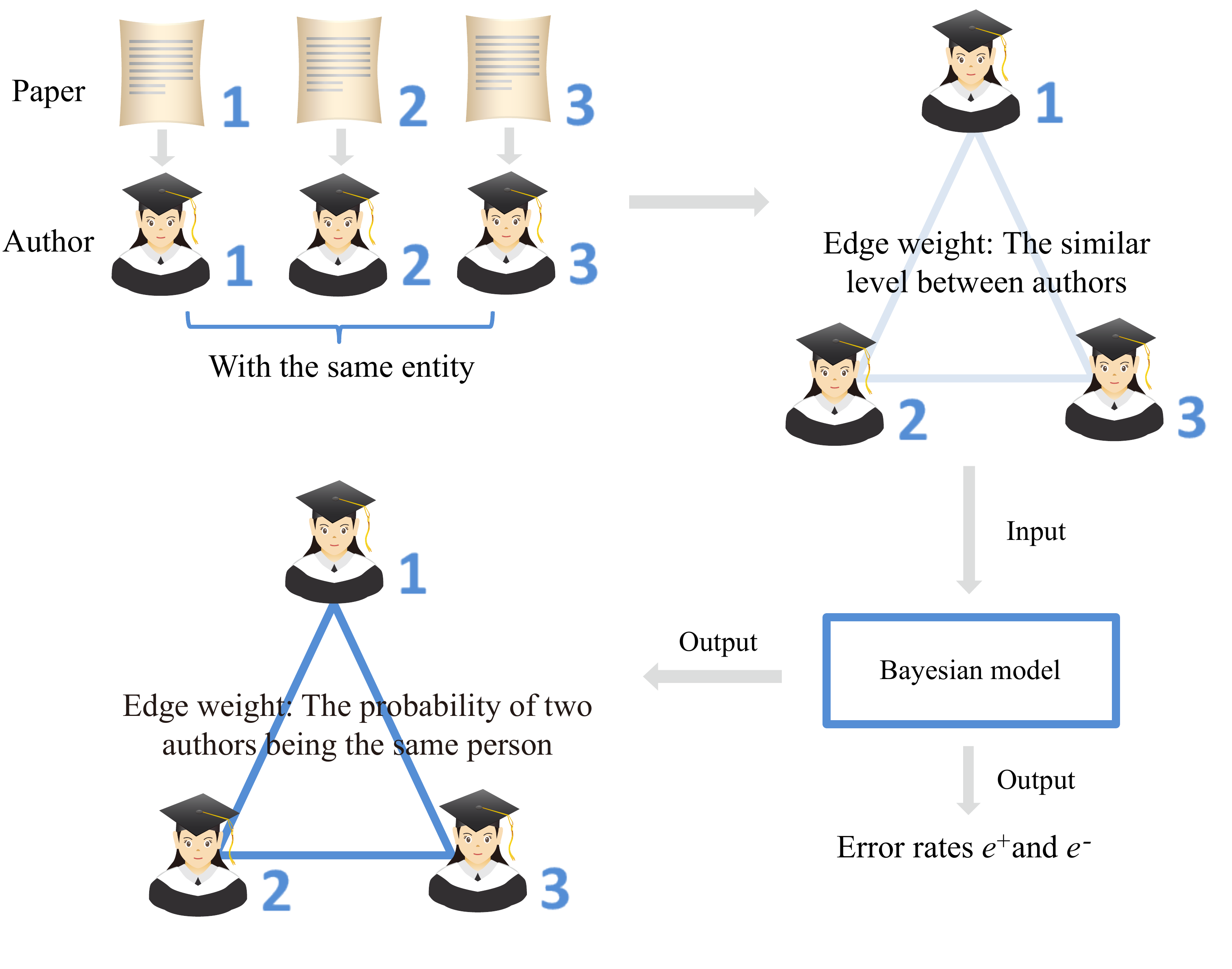}
\caption{    {\bf  Illustration of the Bayesian model.}
The   similar level between   authors is calculated based on author  features, such as   full name,  coauthors, affiliation,
  email address,  etc.
}
 \label{fig1}
\end{figure*}

    Our model
is designed to calculate  the posteriors of  $A_{ij}$ and  $e^\pm$    given   $E_{ij}$ and $N_{ij}$.
Note that its aim is not to  correcting data errors, but to  assessing the merging  error level   of empirical data.
Fig.~\ref{fig1}
shows an illustration of the model.

The unnormalized posterior of $A_{ij}$ given $N_{ij}$, $E_{ij}$,    $e^+$ and $e^-$  is
 \begin{align}\label{eq4} p(A_{ij}|E_{ij},N_{ij},  e^+, e^-) &\propto p(E_{ij},N_{ij}|A_{ij}, e^+, e^-)p(A_{ij})\\\notag
=&
B(E_{ij},N_{ij}|1-e^-)^{A_{ij}} \times B(E_{ij},N_{ij}|e^+) ^{(1-A_{ij}) }\\\notag
&\times \rho^{A_{ij}}(1-\rho)^{1-A_{ij}}
\end{align}
 Eq.~(\ref{eq4}) is   an unnormalized form of a Beta density. Maximizing it with respect to $A_{ij}$
   gives rise to
 \begin{align}\label{eq5}   {A}_{ij} = \frac{\rho (1-e^-)^ {E_{ij}}{e^-}^{N_{ij}-E_{ij}}   }{ \rho (1-e^-)^{ E_{ij}}{e^-}^{N_{ij}-E_{ij}} +(1-\rho) {e^+}^{E_{ij}}(1-e^+)^{N_{ij}-E_{ij}}       }. \end{align}

The same logic  can be utilized  for $e^+$, namely
 \begin{align}\label{eq6} p(e^+|(E_{ij}),(N_{ij}), (A_{ij}), e^-)&\propto p((E_{ij}),(N_{ij})|(A_{ij}), e^+, e^-)p(e^+)
 \notag\\  = &p( e^+) \prod_{i<j} p(E_{ij},N_{ij}|A_{ij},  e^+, e^- )
\notag\\ \propto& \mathrm{Beta}(e^+|\alpha^++1, \beta^++1) \times\prod_{i<j}  B(E_{ij},N_{ij}|e^+)^{1-A_{ij}}
 \notag \\   \propto &   {e^+}^{  \sum_{i<j}E_{ij}(1-A_{ij})  + \alpha^+} \times(1-e^+) ^{ \sum_{i<j}(N_{ij}-E_{ij})(1-A_{ij}) +\beta^+},
   \end{align}
   where $(A_{ij})$, $(E_{ij})$ and $(N_{ij})$ are $N_l\times N_l$ matrixes.
 Maximize it with respect to $e^{+}$ and perform  the same calculation for $e^-$. Then, we  obtained
 \begin{align}\label{eq7} {{e}}^+& =  \frac{ \sum_{i<j}E_{ij}(1-A_{ij})  + \alpha^+}{ \sum_{i<j} N_{ij} (1-A_{ij}) +    \alpha^+ +\beta^+  } ,\notag\\
 {{e}}^- &= 1- \frac{ \sum_{i<j}E_{ij}A_{ij}  + \beta^-}{ \sum_{i<j} N_{ij} A_{ij} +    \alpha^- +\beta^-  } . \end{align}

Value $\alpha^\pm$ and $\beta^\pm$.
Initialize     $e^\pm$  and $\rho$.
 Iterate  Eqs.~(\ref{eq3}), (\ref{eq5}) and~(\ref{eq7}) until   convergence or the  maximum number of iterations is reaching. Then we can obtain
  $({A}_{ij})$,  ${e}^\pm$ and $\rho$. Table~\ref{tab1} shows this process.

\begin{table*}[h] \centering \caption{{\bf The process of the presented    model.} }
\begin{tabular}{l r r r r r r r r r} \hline
Input:  Observed $N_l\times N_l$ matrixes  $(E_{ij})$ and $(N_{ij})$ for entity $l$.\\
\hline
Value    $\alpha^+$,  $\alpha^-$, $\beta^+$  and $\beta^-$;\\
Initialize    $e^+$, $e^-$  and $\rho$;
\\
Repeat:\\
~~~~calculate $A_{ij}$ through Eq.~(\ref{eq5});\\
~~~~renew  $e^+$ and $e^-$ through Eqs.~(\ref{eq7});\\
~~~~renew $\rho$ through Eq.~(\ref{eq3}).\\
Until  convergence or the maximum number of iterations   is reached.\\ \hline
Output:    $(A_{ij})$,   $e^\pm$ and $\rho$. \\ \hline
 \end{tabular}
\label{tab1}
\end{table*}

The values of error rates help us evaluate the contribution  of  the given  features.
 Error rates $e^+\rightarrow\alpha^+/(\alpha^++\beta^+)$ and $e^-\rightarrow\alpha^-/(\alpha^-+\beta^-)$
  imply that the  features, reflected by $E_{ij}$ and $N_{ij}$, carries no information.
 More broadly, the carried information can be measured by
 \begin{align}\label{eq8}I=1 -(e^+ +
e^-).\end{align}
  For example, if  a feature with $E_{ij}=N_{ij}$ for any $i$ and $j$, then
$A_{ij}=1$ is a solution of Eq.~(\ref{eq5}).
Submitting the solution to Eqs.~(\ref{eq7}) gives rise to
${e}^+=\alpha^+/(\alpha^++\beta^+)$, ${e}^-=\alpha^-/({   N_l(N_l-1)/2 +    \alpha^- +\beta^-  } )$, which
  are free of the feature.  It means no  information is carried by the feature.
    If  a feature with $E_{ij}=0$ for any $i$ and $j$, then
$A_{ij}=0$ is a solution of Eq.~(\ref{eq5}), then the error rates
 ${e}^+= \alpha^+/(N_l(N_l-1)/2+\alpha^++\beta^+)$, ${e}^-= \alpha^-/({     \alpha^- +\beta^-  } )$, which are   also free of the feature.


The formula (\ref{eq8}) assesses  the contribution of the given  features  to the improvement of the quality of empirical data.
The contribution here means that using  the feature helps us decrease the uncertainty of estimating the ground-truth data, namely $(A_{ij})$ for any $i$ and $j$.
Increasingly positive or negative value  of $I$ indicates increasing or decreasing  information of the feature\cite{Butts}.
 Based on  the features  with positive information,
Eqs.~(\ref{eq7}) can
 tell us
whether the error level  is severe enough to warrant data reliability, and  which name entities are most heavily compromised.

Eq.~(\ref{eq2})
  tells us  the probability of $N_l$ authors being the same person.
With the probability $\rho$, the estimated number of the persons wrongly merged
 is
  $(1-\rho)( {N}_l -1)$. Adding $1$ to this number  gives rise to the
evaluated number of the persons with   entity $l$
\begin{align}\label{eq9}\hat{N}_l&=(1-\rho) N_l+\rho.  \end{align}
Therefore,  we   obtained the evaluated number of a dataset's persons, namely $\hat{N}  = \sum_{l}  \hat{N}_l$.

\section*{Demonstration of  model functions}

The feature of coauthors  (e.~g. the names of coauthors)  has been adopted by a range of  name disambiguation methods\cite{Hussain}.
Here we used coauthors' name to  demonstrate   model functions   on three  sets of papers:
 SCAD-zbMATH (28,321 mathematical papers   during 1867--1999), PNAS-2007--2015 (36,732
papers of  Proceedings of the National Academy of Sciences), and   PRE-2007--2016 (24,079 papers of   Physical Review E).
  The first  set only contains the papers  for which all authors
are manually disambiguated\cite{Muller}.
The ground-truth entities of authors are denoted as
  author id.
   The other two  sets are collected  from Web of Science (www.webofscience.com), where
their papers are  published during those years in their name.
They    contain    the names of authors on papers.

We applied our model to
the entities generated through author id (zbMATH-a), to the entities generated through surname and
the initials of all   provided given names (PNAS-a, PRE-a), to the
entities generated through the names on papers (zbMATH-b, PNAS-b, PRE-b), and to the entities through
 short name, namely surname and the initial of the first given name  (zbMATH-c, PNAS-c, PRE-c).
 Table~\ref{tab2} shows certain statistical indexes for the nine datasets.
\begin{table}[ h]
   \centering
  \begin{threeparttable}
  \caption{\bf  Specific   statistical   indexes of    empirical data. }
  \begin{tabular}{lccrrrrrrrrrrrrrrr}
    \toprule
   Data   & $N$ &  $M$   & $a$ &  $b$  &   $c$  &  $d$ && \\
  \bottomrule
            \multirow{1}{*}{zbMATH-a}   & 2,946&   \multirow{3}{*}{28,321}  &   4.177  &   11.48   &   \multirow{3}{*}{1.194}  & 0.593
  \cr
              \multirow{1}{*}{zbMATH-b}  & 4,696 & &2.621 &7.200&  & 0.586	\cr
    \multirow{1}{*}{zbMATH-c}  &2,919   && 4.216  &11.58 && 0.661 	\cr
    \bottomrule
            \multirow{1}{*}{PNAS-a}   &136,322  &   \multirow{3}{*}{ 34,630 } & 17.82 &  1.793  & \multirow{3}{*}{7.059}&  0.310
  \cr
              \multirow{1}{*}{PNAS-b}  &  161,780  &  & 15.02&1.511 & &   0.259 	\cr
    \multirow{1}{*}{PNAS-c}    & 115,463  &    	&21.04  &2.117 &  &  0.356 \cr
    \bottomrule
            \multirow{1}{*}{PRE-a}   &  30,552  &   \multirow{3}{*}{ 21,634 } & 7.495   &2.365 &  \multirow{3}{*}{  3.339 } & 0.412
  \cr
              \multirow{1}{*}{PRE-b}  & 36,915 &  &6.203  & 1.957&&	 0.355\cr
    \multirow{1}{*}{PRE-c}  &27,925 &   & 8.201	&2.587   & & 0.443\cr
    \bottomrule
    \end{tabular}
    \label{tab2}
    \end{threeparttable}
  \begin{flushleft}
      Index  $N$: the number of entities, $M$: the number of hyperedges, $a$: the average degree,
 $b$: the average hyperdegree,
  $c$:  the average size of hyperedges,
  $d$: the proportion of the entities with hyperdegree$>1$.
    \end{flushleft}
\end{table}

  Given a similar level  $E_{ij}/ N_{ij}$   between any two merged authors $i$ and $j$, we can
 demonstrate   model functions   on  empirical data.
Let $E_{ij}=1$  if $i$ and $j$ have some coauthors sharing the same  short name,
$E_{ij}=0$ if not,
and  $N_{ij}=1$.
 Note that the matrix  $(E_{ij})$ also suffers merging  errors, so it would not be always helpful for name disambiguation.
Even without merging errors, its contributions would also be
limited, because some   authors would  publish  papers  with different authors.

We run the model with  the two settings in Table~\ref{tab3} for   parameters  ($\alpha^\pm$,  $\beta^\pm$)   and initializations ($e^\pm$, $\rho$).
 Our model with  Setting 2   is just the model of Newman\cite{Newman9}.
  The reason of using Setting 1  is that it satisfies  following  restrictions.
The number of pairs   $\epsilon=N_l(N_l-1)/2$, which gives rise to   $\alpha^++\beta^+=\alpha^-+\beta^-=\epsilon$.
When $E_{ij}=1$ for all possible $i$ and $j$, we let $e^+=1$, $e^-=0$  and $\rho=1$, because all of the $N_l$ authors  are regarded as the same person.
It gives rise to   $\alpha^-=\beta^+=0$.
When $E_{ij}=0$, we  let $e^+=0$, $e^-=1$  and $\rho=0$, because all of the $N_l$ authors  are regarded as different persons. It gives rise to    $\alpha^+=\beta^-=0$.

 We compared the results of our model with those of Newman's model.  The outputs of the two models are listed in
Table~\ref{tab4}. For   zbMATH-a,b,c,
their estimated number of authors  $\hat{N}$ deviates   far away from the number of entities $N$.
  There are 59.3\% authors published more than one paper. The average hyperdegree of authors is   $11.48$.
 Therefore, the size of matrix $(E_{ij})$   for over 59.3\% entities is larger than $11\times 11$.
Meanwhile, the average number of the authors per paper  is only $1.194$, thus   $(E_{ij})$ is   sparse   for many entities.
It gives rise to the low value of $\rho$ for those entities,   and thus leads to the large value of the ratio $\hat{N}/N$.

\begin{table*}[h] \centering \caption{{\bf The settings of  parameters  and initializations.} }
\begin{tabular}{l r r r r r r r r r} \hline
 Setting 1:  $\alpha^+=\beta^-=\epsilon\rho_0$,     $\alpha^-=\beta^+=\epsilon (1-\rho_0)$, $e^+=\rho=\rho_0$ and   $e^-=1-\rho_0$,  \\
  ~~~~~~~~~~~~~where  $\rho_0= \sum_{l,k}E_{lk}/\sum_{l,k}N_{lk}$, and   $\epsilon= {N_l(N_l-1) }/{2}$.\\
\hline
Setting 2: $\alpha^+=\beta^-=\alpha^-=\beta^+=0$,   $e^+$, $e^-$  and $\rho$ are  random variables drawn \\
 ~~~~~~~~~~~~~from  the  uniform distribution $U(0,1)$. \\ \hline
 \end{tabular}
\label{tab3}
\end{table*}


\begin{table}[ h]
   \centering
  \begin{threeparttable}
  \caption{\bf  The comparisons between the outputs of our model   and those of Newman's model. }
    \begin{tabular}{lrrrrrrrrr}
    \toprule
   Data  &   $ \hat{N}$ &  $ \hat{N}/N$ &     $ \bar{{e}}^+ $ & $ \bar{{e}}^- $ &   $ \bar{{I}}$ &  $ \bar{\rho}$ \\
\cmidrule(lr){1-7}
            \multirow{2}{*}{zbMATH-a} &       30,943& 1050.4\% & 0.304&0.696 &0.000&0.304
  \cr
      &    19,072  &  647.4\% &0.313& 0.692&-0.004&0.499
  \cr 
\cmidrule(lr){1-7}
              \multirow{2}{*}{zbMATH-b}  &  30,722 & 654.2\% & 0.247&0.753&0.000&0.247	\cr 
            &   19,036	&405.3\%   &0.255& 0.747 &-0.001 &0.502	\cr  
  \cmidrule(lr){1-7}
    \multirow{2}{*}{zbMATH-c}  &  21,218& 726.9\% & 0.281& 0.719 &0.000& 0.281 \cr  
  &  18,372&629.4\%& 0.288&0.711&0.001&0.504    \cr  
    \bottomrule
            \multirow{2}{*}{PNAS-a}   &    201,699 &148.0\%&0.537 &0.463&0.000&0.537
  \cr  
      &     190,445& 139.7\% & 0.546 &0.455 &-0.001&0.500  \cr  
\cmidrule(lr){1-7}
              \multirow{2}{*}{PNAS-b}    & 198,231   & 122.5\%	& 0.637  &0.363 & 0.000& 0.637 	\cr 
            &  203,420   & 125.7\% & 0.643 &0.357  &	0.001 & 0.500\cr  
  \cmidrule(lr){1-7}
    \multirow{2}{*}{PNAS-c}   & 186,601 &161.6\%&0.469 &0.531&0.000&0.469 \cr   
  &  180,628&156.4\%& 0.479&0.519&0.002&0.498 \cr
    \bottomrule
            \multirow{2}{*}{PRE-a}   &  53,230 & 174.2\% & 0.585 & 0.415 &0.000&0.585
  \cr 
      &   51,387 &168.2\%& 0.594&0.406&0.000&0.501
  \cr 
\cmidrule(lr){1-7}
              \multirow{2}{*}{PRE-b}  & 53,159    & 144.0\% 	&  0.646 &0.354& 0.000	&  0.646\cr   
            &  54,450&  147.5\% 	&  0.653    &0.347&0.001 &0.500	\cr  
  \cmidrule(lr){1-7}
    \multirow{2}{*}{PRE-c} & 50,503 &180.9\%& 0.554 &0.446&0.000& 0.554    \cr  
  &  49,654&177.8\% &0.564&0.438& -0.002& 0.498   \cr  
    \bottomrule
    \end{tabular}
    \label{tab4}
    \end{threeparttable}
  \begin{flushleft}
For each sub-table,  the outputs of our model are listed in the first row, and  those of Newman's model in the second row.
      Index  $\hat{N}$: the evaluated  number of entities, $\hat{N}/N$: the ratio of $\hat{N}$ to the number of entities $N$,
    $ \bar{\rho}$:  the average of entities' $\rho$ defined by Eq.~(\ref{eq2}),   $ \bar{e}^\pm $:
 the average of entities' $  {e}^\pm $ defined by Eqs.~(\ref{eq7}),
and  $ \bar{{I}}$:  the average of entities' $I$ defined by the formula~(\ref{eq8}).
 \end{flushleft}
\end{table}

The ratios  of  PNAS-a,b,c are smaller than those of zbMATH-a,b,c,   and   those of   PRE-a,b,c,  respectively.
 For PNAS-a,b,c,
 their  average hyperdegree
is   relatively small, and
their average
degree is  relatively large.
Therefore,  their $(E_{ij})$ is   dense and has
  a   small  size on average, compared to that of other datasets.
It gives rise to  a small  value  of  their ratio $\hat{N}/N$.

\begin{figure*}[h]
\centering
 \includegraphics[height=3.   in,width=4.5  in,angle=0]{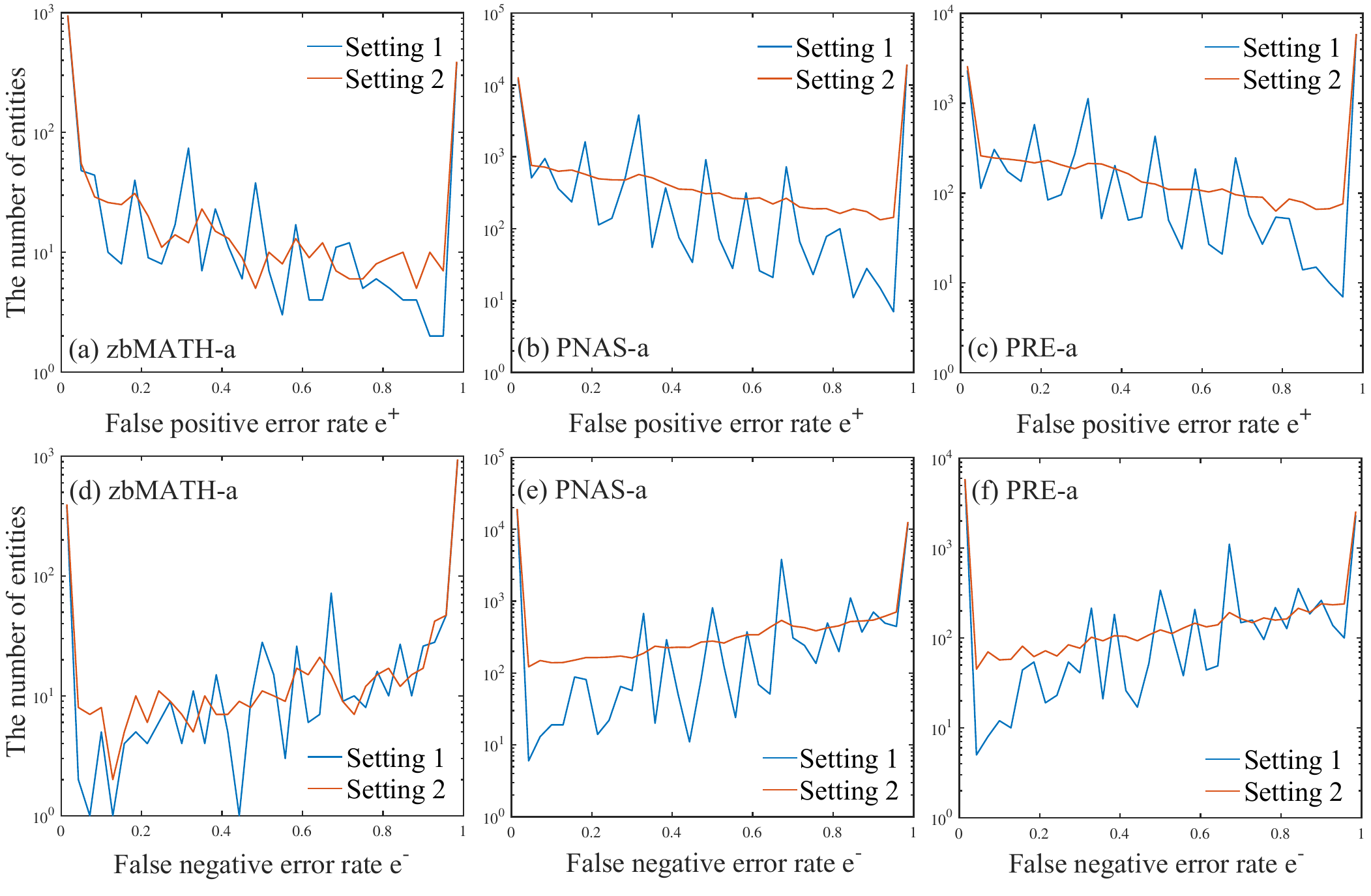}
\caption{    {\bf The distributions  of  error rates.}
Panels show the comparisons between
the distributions
calculated through our model (Setting 1)
  and those through Newman's model (Setting 2).
}
 \label{fig2}
\end{figure*}

Fig.~\ref{fig2} shows  the distributions of false positive rates and those of false negative rates
 for the datasets with the suffix -a.
We found that the trends of those distributions  calculated through our model and those through Newman's  model   are the same.
The number of entities with   error rate $1$ or $0$ is   large, compared with   that of any other value.
The distributions of false positive rates are symmetric to the corresponding distributions of false negative rates about   $0.5$.
This is because  that
coauthors' short name carries
  null information, namely $e^++e^-\approx1$.

\begin{figure*}[h]
\centering
 \includegraphics[height=6.   in,width=4.5  in,angle=0]{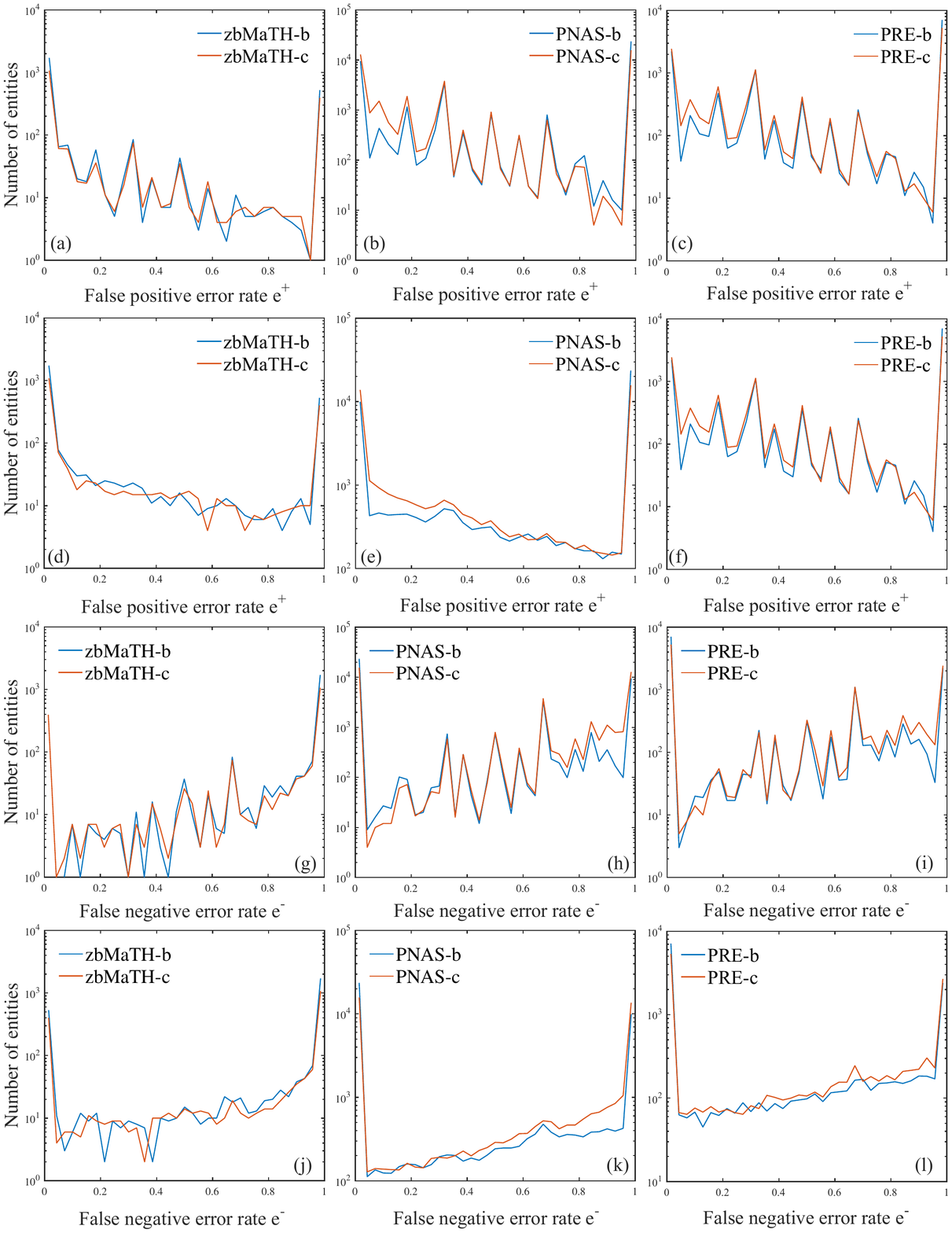}
\caption{    {\bf The error distributions of the entities generated through different methods.}
 The entities of the datasets with suffix -b are generated through the names on papers, and those with suffix -c are generated through short name. The distributions in Panels (a-c, g-i) are calculated through our model, and those in Panels (d-f, j-l) are calculated through Newman's model.
 }
 \label{fig3}
\end{figure*}

 Fig.~\ref{fig3} shows these distributions  for  the datasets with suffixes -b and -c.
 We found that these distributions for the entities obtained by different methods are     positively correlated. In fact,
 the Pearson correlation coefficient of   each pair from the datasets with the same prefix  is larger than $0.95$.  However,
the error rates are different for the entities obtained through different methods.
The highly positive correlations imply that the feature of coauthors' short name is
insensitive to merging errors.

\begin{figure*}[h]
\centering
 \includegraphics[height=5.6  in,width=4.5  in,angle=0]{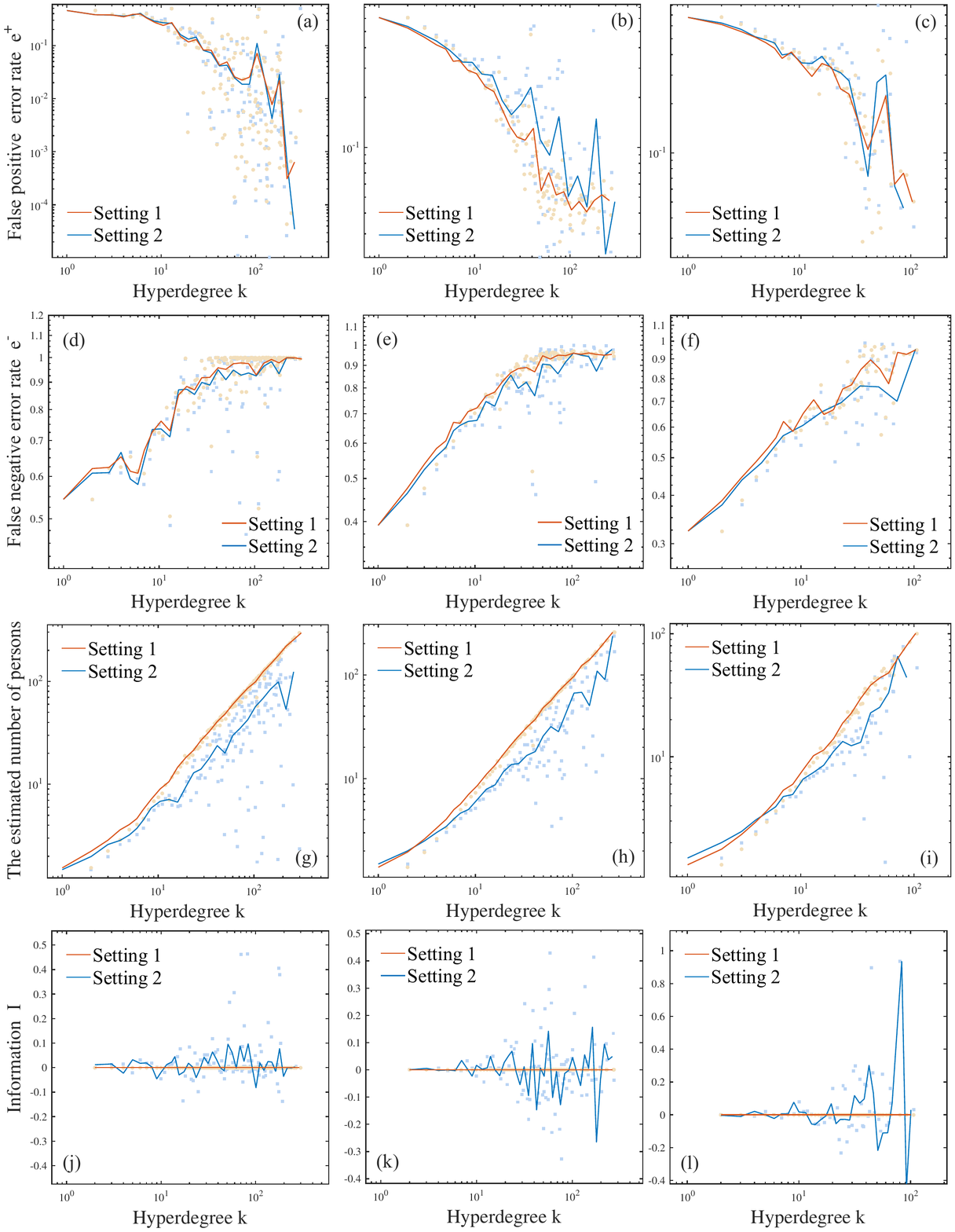}
\caption{    {\bf The relationship between hyperdegree  and  model outputs.}
The inputs of models are the datasets  with suffix-a.  Panels show the   average value of each output (i.~e., $e^+$, $e^-$,  $\hat{N}_l $ and $I$) over the  entities  with hyperdegree $k$ (red circles, blue squares).
 These average values are binned
on abscissa axes  to     extract the trends
   hiding in   noise (red and blue lines). }
 \label{fig4}
\end{figure*}

Because of the positive correlations, we now only discuss the model outputs for the datasets with suffix -a.
 Fig~\ref{fig4} shows the relationship between the model outputs ($I$, $\hat{N}$, and ${e}^\pm$) and hyperdegree. We found that the average value of the false positive rates over $k-$hyperdegree entities decreases with the growth of $k$. The case for the false negative error rates and that for the estimated numbers of persons  are reverse. Underlying these phenomena is that the average value of $A_{ij}$ over $k-$hyperdegree entities decreases with the growth of $k$. This is caused by the property of empirical data: the probability that the authors of an entity have a coauthor with the same short name decreases with the growth of the entity's hyperdegree, on average.

  These phenomena shown in Fig~\ref{fig4}  mean
  disambiguating by using   the
 short names (surname and the initial of the first given name) of coauthors
 surfers high risk of false positive  errors but low risk of false negative errors
for the entities with a small hyperdegree.  The case is reverse for the entities with a large hyperdegree.
 However, Fig~\ref{fig4}   also shows  that
the  average value of the information  $I$ over $k-$hyperdegree entities is around   $0$ for any possible  $k$.  It means that  synthetically considering both risks, the
 short names  of coauthors carry null information for improving the  quality of the empirical datasets.

\section*{Discussion and conclusions  }

A Bayesian model is provided to measure the quality of coauthorship data based on author features, and to assess the contribution of author features to the improvement of data quality. With the model, we can calculate the distribution of merging errors   over the entities of authors, inferring the number of persons. Even where our model does not provide substantial uncertainty reduction (e.~g., where only the information of coauthors' name is available), it may still be of use in its ability to provide a theoretical assessment of merging error level for coauthorship data.

Network scale and data accuracy are two irreconcilable challenges for coauthorship analysis. It is easy to correct a small dataset, but the results obtained by studying small datasets are incomplete. Meanwhile, correcting a large dataset is a time-consuming mission. Even corrected data with large scale are available, they still cannot cover all coauthorship. Many existing results of coauthorship are based on incomplete or imperfect coauthorship data\cite{Xie3,Xie7,Xie6}. Our model has the potential to be extended for assessing the confidence level of these results, thus would have clear applicability to empirical research.

\section*{Funding}
ZX acknowledges support from National Science Foundation of China (NSFC) Grant No. 61773020.

\section*{Acknowledgments}
The author thinks
Miao Li in the KU Leuven,   and Jianping Li in the National University of Defense Technology for their helpful comments and feedback.

\end{document}